\documentclass{article}

\usepackage{arxiv}
\usepackage[utf8]{inputenc} 
\usepackage[T1]{fontenc}    
\usepackage{hyperref}       
\usepackage{url}            
\usepackage{booktabs}       
\usepackage{amsfonts}       
\usepackage{nicefrac}       
\usepackage{microtype}
\usepackage{lipsum}		
\usepackage{graphicx}
\usepackage{doi}
\usepackage{listings}
\usepackage[font=normalsize,labelfont={bf}]{caption}

\hypersetup{
    colorlinks=true,
    linkcolor=blue,
    urlcolor=blue,
}

\graphicspath{ {./images/} }

\title{Improving MATLAB's \emph{isprime} performance without arbitrary-precision arithmetic}

\date{August 1, 2021}	

\author{ {\hspace{1mm}Travis Near} \\
	Department of Computer Science \\
	Missouri University of Science and Technology \\
	Rolla, MO 65409 \\
	\texttt{tentxb@mst.edu} \\
}

\renewcommand{\headeright}{}
\renewcommand{\undertitle}{}

\begin{document}
\maketitle

\begin{abstract}
MATLAB is a numerical computing platform used by scientists, engineers, mathematicians, and students which contains many mathematical functions, including \emph{isprime}. MATLAB's \emph{isprime} function determines which elements of an input array are prime. This research details modular arithmetic techniques, the Miller—Rabin primality test, vectorized operations, and division-minimizing strategies which harness the power of MATLAB's capabilities to improve \emph{isprime}'s performance. The results are typically 5 to 10 times faster for small integers and many hundreds of times faster for large integers and long arrays.
\end{abstract}


\section{Introduction}
MATLAB's \href{https://www.mathworks.com/help/matlab/ref/isprime.html}{\emph{isprime}} function accepts an input array, \emph{N}, then returns which elements in the array are prime:

\begin{verbatim}
    >> isprime([1, 2, 3, 4, 5])
    ans =

      1x10 logical array

      0   1   1   0   1 
\end{verbatim}

\emph{isprime} begins with the sieve of Eratosthenes followed by trial division of all primes up to the square root of the maximum value, \emph{M}, in the array \cite{isprime}. The sieve begins with an incrementing sequence of \([2, 3, 4, 5, ..., \emph{M}]\). Then it \emph{crosses-off} multiples of primes. Two, for example, is prime so all multiples of two are crossed-off given that they share a common factor. This leaves \([2, 3, 5, 7, 9, ..., \emph{M}]\). The next prime is three: it is marked prime then its multiples are crossed-off leaving \([2, 3, 5, 7, 11, ..., \emph{M}]\). The process repeats until all prime numbers have been visited.

Next, \emph{isprime} utilizes trial division of all primes returned by the sieve. \emph{isprime} divides each element of the input array by all primes which are \emph{less} than that element. The input element is marked prime when none of the smaller primes divides into it evenly. This algorithm is simple, effective, and accurate, but is not optimized for any category of inputs. 

This research introduces \emph{isprime\_fast}, a replacement for \emph{isprime}. \emph{isprime\_fast}'s inputs and outputs are identical; the only difference lies in the techniques used to determine primality. \emph{isprime\_fast} specializes for four classifications of inputs: (1) large scalar integers, (2) medium scalar integers, (3) small scalar integers, and (4) arrays.

\section{\texorpdfstring{Improving the performance for large scalar integers (2\textsuperscript{49} to 2\textsuperscript{64})}{Improving the performance for large scalar integers}}
\label{sec:headings}

MATLAB's \emph{isprime} takes about 30 seconds to check primality for 64-bit integers. \emph{isprime\_fast} reduces this to about 0.1 seconds by implementing a modified Miller–Rabin primality test. MATLAB does not support arbitrary precision arithmetic, so the modifications made to Miller–Rabin use modular arithmetic properties to clamp the range of intermediate values to ensure a correct result.

Miller–Rabin is a \emph{probabilistic} prime number algorithm, yet \emph{isprime} is \emph{deterministic}. To replace \emph{isprime}, \emph{isprime\_fast} must use a deterministic version of Miller–Rabin. The reason that Miller–Rabin is probabilistic is due to its selection of \emph{bases} to check congruence relations \cite{Rabin}. To produce deterministic results, Miller–Rabin needs to select bases which produce the correct result for \emph{isprime}'s entire input domain of 0 to 2\textsuperscript{64}.

\subsection{The Miller–Rabin primality test}
Miller–Rabin is built on two properties which hold for all prime numbers. First, Fermat's little theorem \cite{Rabin} which states that for an integer base \(a\) and prime number \(n\),
\begin{equation} \label{eq:1}
    a^{n-1} \equiv \pm 1 \pmod{n}
\end{equation}
Second, the square roots of Equation \ref{eq:1} also hold. By rewriting \(n-1\) as powers of two multiplied by a constant \(d\), this becomes,
\begin{equation}
    a^{2^{s}d} \equiv \pm 1
\end{equation}

By successively removing powers of two, the terms to check for congruence are,
\begin{equation}
a^{2^{s}d}, a^{2^{s-1} d}, a^{2^{s-2} d}, ..., a^d
\end{equation}
If \(n\) is prime, all elements in this sequence are congruent to \(\pm 1\) (mod \(n\)) \cite{Lynn}.

These two properties hold for all prime numbers. However, there exist composite numbers that pass these two tests which are known as \emph{strong pseudoprimes}. The likelihood that a number is truly prime increases as more bases are checked against the two properties above. By carefully selecting bases, Miller–Rabin can be made deterministic up to any arbitrary value. To replace \emph{isprime}, Miller–Rabin must use bases which return the correct answer for inputs up to 2\textsuperscript{64}. The following bases provide accurate results for all 64-bit integers \cite{Sinclair}:
\begin{equation}
    2, 325, 9375, 28178, 450775, 9780504, 1795265022
\end{equation}

\emph{isprime\_fast} implements the Miller–Rabin algorithm as detailed by its original authors. However, it makes several modifications to ensure that no intermediate result at any point exceeds MATLAB's maximum integer value of 2\textsuperscript{64}. This adapted Miller–Rabin algorithm is built on three core functions: \emph{ModAdd}, \emph{ModMultiply}, and \emph{ModExp}.

\subsection{ModAdd}
\emph{ModAdd} calculates \(c = a + b \pmod{m}\). This function adds two unsigned 64-bit integers followed by (\emph{mod m}) without overflow. The temporary result of \(a + b\) (before \emph{mod m}) can overflow (more specifically, \emph{saturate}) and return the highest possible value given the precision available. For example, \(100 + 200 = 300\) but for unsigned 8-bit integers (\emph{uint8}) the result is the largest value allowed by this data type: \(100 + 200 = 255\) for \emph{uint8} in MATLAB. Saturation is undesirable for \emph{isprime\_fast} because it surrenders accuracy. To avoid overflowing, \emph{ModAdd} uses modular arithmetic properties, invariants, and careful ordering of the sequence of its calculations:

\begin{itemize}
    \item \(c = a + b \pmod{m} \) \emph{[intended calculation]}
    \item \(assert(a + b > m)\)
    \item \(assert(m \geq b)\)
    \item \(assert(m \geq a)\)
    \item \(assert(m \leq 2^{64})\)
    \item Given that \(a + b > m\), the inequality \(a > m - b\) must hold
    \item Given that \(m \geq b\), the result \(m - b\) must be positive or zero
    \item \(c = a + b - m\) \emph{[mod can be simplified to subtraction given that a and b are less than m]}
    \item \(c = a - (m - b)\) \emph{\textbf{[Property 1]}}
\end{itemize}

\subsection{ModMultiply}
\emph{ModMultiply} calculates \(c = a * b \pmod{m}\). This function multiplies two unsigned 64-bit integers followed by \emph{mod m} without overflow. Like \emph{ModAdd}, the temporary result of \(a * b\) before \emph{mod m} will overflow when \(a * b\) exceeds the range of 64-bit integers. An optimized implementation of \emph{ModMultiply} considers four scenarios:

\begin{enumerate}
    \item Fast path: \(a * b\) does not overflow
    \begin{enumerate}
        \item \(c = a * b \pmod{m}\) \emph{[intended calculation]}
    \end{enumerate}
    \item \(a = b\)
    \begin{enumerate}
        \item \(c = a * b \pmod{m}\)
        \item \(assert(m \geq a)\)
        \item \(c = a * a \pmod{m}\) \emph{[substitute a=b]}
        \item \(c = (a \pmod{m} * a \pmod{m}) \pmod{m}\) \emph{[modular distribution]}
        \item \(c = (m - a) * (m - a) \pmod{m}\) \emph{[mod is based on difference between m and a]}
        \item \(c = (a * a) \pmod{m} = (m - a) * (m - a) \pmod{m}\)
        \item \(aSmall = min(m, m - a)\) \emph{[from line f, m and m – a give the same result]}
        \item if \(aSmall \leq 2^{32}\); \(c\) can be calculated without overflow using \emph{uint64}, \(c = aSmall^2 \pmod{m}\)
        \item otherwise, proceed to case 4 (general overflow)
    \end{enumerate}
    \item \(a = 2\)
    \begin{enumerate}
        \item \(c = a * b \pmod{m}\) \emph{[intended calculation]}
        \item \(assert(m \geq b)\)
        \item \(assert(2 * b > m)\) \emph{[must be true given that \(a * b > 2^{64} > m\)]}
        \item \(c = 2 * b \pmod{m}\) \emph{[substitute a=2]}
        \item \(c = b + b \pmod{m}\) \emph{[replace multiplication with addition]}
        \item \(c = b - (m - b)\) \emph{[by Property 1]}
    \end{enumerate}
    \item \(a * b\) overflows 64-bit integers
    \begin{enumerate}
        \item A common modular-multiply implementation \cite{GeeksforGeeks} (shown below as \emph{ModMultiplyClassic}) computes \(a * b \pmod{m}\) by reducing to \(mod(2 * a, m) * (b / 2)\). 
        \item This increases \emph{ModMultiply}'s range, but still encounters an overflow when \(2^{63} < a \leq 2^{64}\). This is due to doubling \(a\) before applying \emph{mod}.
        \item To replace \emph{isprime}, \emph{ModMultiply}'s range must be doubled to reach \(2^{64} (2 * 2^{63})\).
    \end{enumerate}
\end{enumerate}

\lstset
{
    language=MATLAB,
    numbers=left,
}

\begin{lstlisting}
function c = ModMultiplyClassic(a, b, m)
    % c = a * b (mod m)
    c = 0;
    a = mod(a, m);
    while b > 0
        if bitand(b, 1) % odd numbers
            c = mod(c + a, m);
        end
        a = mod(bitshift(a, 1), m);
        b = bitshift(b, -1);
    end
    c = mod(c, m);
end
\end{lstlisting}

Lines 7 and 9 are where this algorithm can overflow for 64-bit integers. Line 7 adds \(c + a\) without overflow guards. This is fixed by \emph{ModAdd}: \(c = ModAdd(c, a, m)\). Line 9 utilizes a \emph{bitshift} of \(a\). Bitshift by one shifts all bits to the left by one, effectively doubling a value. For values ranging from \(2^{63}\) to \(2^{64}\), a naive bitshift loses the highest bit and the result \emph{decreases} rather than doubles. \emph{ModAdd} can resolve this overflow as well, \(a = ModAdd(a, a, m)\).

These two updates fix \emph{ModMulitplyClassic}'s overflow issues. This implementation now works correctly for all 64-bit numbers by avoiding saturating overflows. However, this revised implementation is not yet optimized for performance due to its large \emph{while} loop and absence of vectorized MATLAB instructions.

\emph{Vectorization} in MATLAB is the operation on multiple elements of an array in one calculation \cite{Vectorization}. For example,
\begin{verbatim}
    >> [1, 2, 3] + 1
    ans =

         2     3     4
\end{verbatim}
The scalar \(1\) implicitly expands into an array then is added element-wise to \([1, 2, 3]\). Vectorization can greatly improve performance due of the memory layout of MATLAB variables: the \emph{data} portion of variables in MATLAB occupy a contiguous memory buffer \cite{MATLABData}. Vectorized operations operate on this entire chunk of memory at once. Contiguous memory is optimized for prefetching, data locality, and branchless execution leading to faster performance \cite{Drepper}.

\emph{ModMultiplyClassic} can be improved by vectorizing its independent calculations. Note that every \(b\) value in \emph{ModMultiplyClassic} is independent of all other calculations: \(b\) is composed of bitshifts followed by checking individual bits. These operations can be lifted out of the \emph{while} loop and vectorized into:
\begin{verbatim}
shift = bitshift(b, 0:-1:-63);
oddIdx = find(mod(shift, 2) == 1);
\end{verbatim}

All 64 bitshifts are done in one statement then vectorized division by two is performed for this array of results to determine which elements are odd. By removing \(b\) from the \emph{while} loop, the loop becomes simpler, faster, and easier to optimize further.

The final version of \emph{ModMultiply} contains one additional vectorized instruction: accumulating the sequence of all intermediate \(a\) values created in the loop. By storing the entire sequence into an array of length 64, each element can be added using MATLAB's vectorized \emph{sum} function: \(c = sum(sequenceA)\). The performance gain for this technique regularly outweighs the cost to allocate 512 bytes (64 elements * 8 bytes = 512) of memory.

\subsection{ModExp}
\emph{ModExp} (Modular Exponentiation) calculates \(c = a^b \pmod{m}\) without overflow. The bulk of its work is handled by the two building blocks \emph{ModAdd} and \emph{ModMultiply}. \emph{ModExp} uses exponentiation by squaring to calculate the result in logarithmic time \cite{Schneier}.

\subsection{Summarizing the Miller–Rabin primality test}
\emph{ModAdd}, \emph{ModMultiply}, and \emph{ModExp} combine to implement the Miller–Rabin test without arbitrary-precision arithmetic. Miller–Rabin is a modern algorithm, utilizing powerful prime number congruence relations with a careful selection of bases to determine primality deterministically. Using these tools, \emph{isprime\_fast} decidedly beats the speed of \emph{isprime} for large integers: for inputs in the 2\textsuperscript{49} to 2\textsuperscript{64} range, \emph{isprime\_fast} ranges from 5x to 2950x faster than \emph{isprime} (results summarized in Fig. \ref{fig:Comparison}).

\section{\texorpdfstring{Improving the performance for medium scalar integers (2\textsuperscript{18} to 2\textsuperscript{49})}{Improving the performance for medium scalar integers}}

MATLAB's \emph{isprime} performs well in this range, however a few helpful techniques make it run 3-8x faster for primes and 7-30x faster for composites.

Profiling \emph{isprime(N)} reveals that most of its execution time is spent sieving for primes \(\leq\) \emph{sqrt(N)}. To improve \emph{isprime}, \emph{isprime\_fast} eliminates the sieve of Eratosthenes completely in favor of a simpler wheel sieve. A wheel sieve is an increasing sequence of numbers with select small primes removed. For example, a 2-wheel is \([1, 3, 5, 7, 9, ...]\), and a 2/3-wheel is \([1, 5, 7, 11, 13, ...]\). After creating this sequence, vectorized trial division is run against the input scalar \(N\). If none of the divisions divides evenly, \(N\) must be prime. Efficiently generating a wheel sieve requires understanding two properties of the wheel:
\begin{enumerate}
    \item The cyclical \emph{differences} between elements of the wheel
    \item The \emph{cycle length}
\end{enumerate}

For example, a 2/3-wheel begins with \([1, 5, 7, 11, 13, ...]\). The difference between consecutive elements repeats: \([4, 2, 4, 2, ...]\). Its \emph{cycle length} equals the product of the removed primes: \(2*3 = 6\). Thus, a 2/3-wheel has a difference cycle of \([4, 2]\) which repeats every 6 numbers. Similarly, a 2/3/5-wheel has a difference cycle of \([6, 4, 2, 4, 2, 4, 6, 2]\) which repeats every \(2*3*5 = 30\) numbers. And the 2/3/5/7-wheel used by \emph{isprime\_fast} has a repeating difference sequence of:
\begin{equation}
    [10, 2, 4, 2, 4, 6, 2, 6, 4, 2, 4, 6, 6, 2, 6, 4, 2, 6, 4, 6, 8, 4, 2, 4, 2, 4, 8, 6, 4, 6, 2, 4, 6, 2, 6, 6, 4, 2, 4, 6, 2, 6, 4, 2, 4, 2, 10, 2]
\end{equation}

which repeats every \(2*3*5*7 = 210\) numbers.

By knowing the sequence of differences and the cycle length, \emph{isprime\_fast} can efficiently replicate the repeating sequence up to the length of \emph{sqrt(N)}. Then, to get the integers needed for trial division, the elements are \emph{cumulatively summed} (\emph{cumsum}). This is when the total sum accumulates in each successive element. For example, to obtain the first six elements needed for division in a 2/3/5/7-wheel, compute:
\begin{equation}
    1 + cumsum([10, 2, 4, 2, 4, 6]) = 1 + [10, 12, 16, 18, 22, 28] = [11, 13, 17, 19, 23, 29]
\end{equation}

Cumulative sum of a repeating sequence is several times faster than \emph{isprime}'s sieve for two primary reasons. First, MATLAB excels at creating sequences using its built-in vectorized functions such as \emph{cumsum} and \emph{repmat} (\emph{rep}eat \emph{mat}rix). Second, creating the sequence requires only one \emph{backtrack} (i.e., visiting a previously visited element). By comparison, the sieve of Eratosthenes requires extra backtracks to cross-off all composite numbers. This leads to its time complexity of \emph{O(N log log N)} \cite{Sedgewick} with comparatively high constant factors compared to a simpler wheel. Creating a wheel in MATLAB is a \emph{linear O(N)} operation which requires only two visits of each element:
\begin{enumerate}
    \item The cycle sequence is repeated up to length \emph{sqrt(N)}
    \item These elements are added in a cumulative sum of length \emph{sqrt(N)}
\end{enumerate}

\emph{isprime\_fast} removes the bottleneck prime number sieve and focuses on efficient, vectorized operations. Its speedup factor for medium size scalar inputs is typically 3x to 30x faster than \emph{isprime} (results summarized in Table \ref{table:PrimePerformance} and Fig. \ref{fig:Comparison}). 

\section{\texorpdfstring{Improving the performance for small scalar integers (0 to 2\textsuperscript{18})}{Improving the performance for small scalar integers}}

For small prime scalars, \emph{isprime\_fast} uses a 2-wheel (i.e., odd numbers) instead of a more elaborate 2/3/5/7-wheel. Inputs in this range are so small that the cost of repeating then summing the sequence of differences becomes significant. Odd numbers have a constant difference between elements. Therefore, creating a sequence of odd numbers requires only a single visit of each element. By checking divisibility against odd numbers using integer division, \emph{isprime\_fast} typically beats \emph{isprime} by a factor of 5 to 10 for scalar inputs in this range. Section \ref{Division} further emphasizes the importance of integer division.

\section{Symbolic Math Toolbox}
The Symbolic Math Toolbox is a supplemental MATLAB product which includes a suite of mathematical functions for equation solving, factoring, prime numbers, and more. The toolbox supports arbitrary-precision arithmetic and is not included with MATLAB. Both \emph{isprime} and \emph{isprime\_fast} have no dependencies on it or any other arbitrary-precision calculations.

For determining prime, the Symbolic Math Toolbox includes a function called \emph{sym/isprime} which utilizes a third-party C library and yields excellent performance for large integers (greater than about 2\textsuperscript{50}). For small integers and long arrays, however, most of the execution time is spent converting data recognizable by one programming interface into the \emph{same} data recognizable by a \emph{different} programming interface. 

\emph{isprime\_fast} is intended to replace MATLAB's \emph{isprime} and \emph{not} \emph{sym/isprime}. For completeness, the performance of all three (\emph{isprime\_fast}, \emph{isprime}, and \emph{sym/isprime}) is compared.

\section{Improving \emph{isprime}'s performance for non-scalar inputs}
Until now, only \emph{scalar} (1x1) inputs have been considered. MATLAB's \emph{isprime} function also supports input arrays, matrices, and higher dimensional inputs. Recall that MATLAB's \emph{isprime(N)} implementation consists of two parts:
\begin{enumerate}
    \item Sieve. The sieve returns all prime numbers up to the square root of the maximum input of the array.
    \item Trial division of primes. \emph{isprime} divides each element of input \emph{N} by all primes which are less than that element.
\end{enumerate}

These two steps have a trade-off: sieving is kept efficient by calculating primes up to \(sqrt(max(N))\) instead of \(max(N)\). However, the second step requires many divisions of those primes to determine primality. Consider these two steps for \emph{isprime(1:10)}:
\begin{enumerate}
    \item Sieve. \emph{floor(sqrt(10))} is 3, so the primes to be checked against for trial division are those \(\leq 3\), which are \([2, 3]\).
    \item Trial division. Dividing element-wise from 1 to 10,
    
    \begin{enumerate}
        \item 1 is greater than neither 2 nor 3, so there are 0 divisions
        \item 2 is greater than neither 2 nor 3, so there are 0 divisions
        \item 3 is greater than 2 but not 3, so there is 1 division (3 divided by 2)
        \item The elements \([4, 5, 6, 7, 8, 9, 10]\) are greater than both 2 and 3, so each of these 7 elements is divided by both 2 and 3. There are 2 divisions * 7 elements = 14 total divisions
    \end{enumerate}
\end{enumerate}

Thus, \emph{isprime(1:10)} performs \(0 + 0 + 1 + 14 = 15\) total divisions. The number of divisions required for incrementing sequences \([1, 2, 3, ..., \emph{N}]\) can be calculated by following MATLAB source. The results are in Table \ref{table:NumDiv}.

\begin{verbatim}
p = primes(floor(sqrt(N)));
numDivisions = sum(arrayfun(@(n) nnz(p < n), 1:N))
\end{verbatim}

\begin{table}[ht]
\centering
\begin{tabular}{lll}
Input (1:N) & Number of divisions & Number of divisions / N \\ \hline
1:10        & 15                  & 1.5                     \\
1:100       & 383                 & 3.8                     \\
1:1000      & 10,840              & 10.8                    \\
1:10000     & 248,940             & 24.8                    \\
1:100000    & 6,490,794           & 64.9                    \\
1:1000000   & 167,923,873         & 167.9                   \\
1:10000000  & 4,459,357,131       & 445.9                   \\
1:100000000 & 122,894,263,604     & 1228.9                 
\end{tabular}
\caption{Number of divisions that \emph{isprime} requires for incrementing sequences.}
\label{table:NumDiv}
\end{table}

\subsection{Why the total number of divisions is important} \label{Division}
Floating-point division is a comparatively slow mathematical operation on CPUs, often 10 times slower than multiplication \cite{Hare}. Demonstrated by Table \ref{table:NumDiv}, the total number of divisions \emph{accelerates} with longer incrementing sequences. Therefore, \emph{isprime}'s performance degrades at an increasing rate for these larger data sets, leaving enormous room for improvement.

Instrumenting \emph{isprime} for the array of 1 to 1,000,000 indicates that \emph{isprime} spends 99.9\% of its duration performing remainder division. To improve this, \emph{isprime\_fast} uses three techniques:

\begin{enumerate}
    \item Improving the speed of division
    \item Shrinking the input array
    \item Minimizing the number of divisions
\end{enumerate}

\subsubsection{Improving the speed of division}
The default and most common data type in MATLAB is double \cite{MATLABData}, so \emph{isprime} performs floating-point division for most inputs. Floating-point division preserves extra decimal precision which is not needed given that primality only considers whole numbers. A simple update to use integer division significantly improves \emph{isprime}'s performance:

\begin{table}[ht]
\centering
\begin{tabular}{ll}
Calculation              & Duration (sec) \\ \hline
isprime(1:1000000)       & 8.702          \\
isprimeIntDiv(1:1000000) & 1.377         
\end{tabular}
\caption{Comparison of floating-point division vs. integer division.}
\label{table:IntDiv}
\end{table}

This 6.3x speedup requires only a simple data cast to \emph{int32}. \emph{isprimeIntDiv} is otherwise identical to \emph{isprime}.

\subsubsection{Shrinking the input array}
One useful optimization is to remove multiples of small primes from the input array before additional processing. For sequences which increment by one, this significantly reduces the size of the input array. Removing elements divisible by 2 shrinks the array size by 50\%, removing 2/3 shrinks by 66\%, 2/3/5 shrinks by 73\%, and 2/3/5/7 shrinks by 77\% \cite{Prime, Primorial}. \emph{isprime\_fast} trims \([2, 3, 5, 7]\) which works well based on heuristics given that removing each additional prime has diminishing returns.

\subsubsection{Minimizing the number of divisions}
Instead of division, a strategy which works well for many long arrays is to calculate \emph{all} primes up to \emph{N} rather than the \emph{sqrt(N)}. Having all primes avoids division completely: \emph{isprime\_fast} simply needs to check for existence of each element of the input array in the set of all primes. MATLAB's prime number sieve returns a \emph{sorted} list of primes, so \emph{binary search} efficiently finds which inputs are prime. MATLAB's versatile \emph{ismember} function is suited for this.

Table \ref{table:IncrementingSequence} compares the duration of incrementing sequences for \emph{isprime}, \emph{isprimeIntDiv}, and the final \emph{isprime\_fast} which includes all three optimizations.

\begin{table}[ht]
\centering
\begin{tabular}{lllll}
Input       & \emph{isprime} (sec) & \emph{isprimeIntDiv} (sec) & \emph{isprime\_fast} (sec) & \emph{isprime\_fast} speedup \\ \hline
1:10000     & 0.0185               & 0.0101                     & 0.000829                   & 22.3x                                \\
1:100000    & 0.3352               & 0.1174                     & 0.005502                   & 60.9x                                \\
1:1000000   & 8.70                 & 1.794                      & 0.05697                    & 152.7x                               \\
1:10000000  & 139.50               & 20.86                      & 0.6294                     & 221.6x                               \\
1:100000000 & 2898.13              & 820.86                     & 6.728                      & 430.7x                              
\end{tabular}
\caption{Performance of checking primality of incrementing sequences.}
\label{table:IncrementingSequence}
\end{table}

The performance for \emph{isprime\_fast} scales in an approximately linearithmic manner, \emph{O(N lg N)}, due to its binary search (\emph{lg N}) of each of the \emph{N} elements. \emph{isprime} slows much more rapidly with larger input sizes due to the accelerating number of divisions that it requires (see Table \ref{table:NumDiv}).

However, incrementing sequences represent only a tiny sample of all possible input arrays. The open question is: when to choose the \emph{sqrt} path and when to choose the \emph{binary search} path?

\subsection{Reconciling the \emph{sqrt(N)} path with the \emph{binary search} path}
There are two possible execution flows for arrays in \emph{isprime\_fast} to be compared:
\begin{enumerate}
    \item Sieving primes up to \emph{sqrt(N)} then integer division of primes.
    \item Sieving primes up to \emph{N} then binary searching for primes.
\end{enumerate}

The decision of which path to take is determined through heuristics. Data tested includes random 32-bit primes, random numbers, random odd numbers, incrementing sequences, odd numbers in a restricted range, odd numbers with small primes removed, pseudoprimes, random tiny and small numbers, data with a normal distribution, and more. Based on this assortment of data, the two biggest influences on performance are (1) the number of elements in the array, and (2) the maximum value within the array. To determine the best fit, arrays were constructed which have roughly \emph{equal} duration for each of the two paths.

The raw performance data is noisy. Due to this volatility, \emph{isprime\_fast} uses a linear best fit to avoid overfitting. With a first-degree polynomial fit, the raw data's R\textsuperscript{2} value is low, \(0.694/1\). Additionally, the \emph{binary search} path usage is niche: it performs superlatively for long arrays with a relatively small maximum value but degrades quickly if misused on shorter arrays with a high maximum value. Conversely, the \emph{sqrt} path (which is what MATLAB's \emph{isprime} always uses) performs adequately for all varieties of inputs. Therefore, from the raw data, only the data points with comparatively low maximum values were selected for the final curve fitting. This errors on the side of caution: binary search is used only when the input data metrics abundantly work in its favor.

\begin{figure}[ht]
    \centering
    \includegraphics[width=11cm]{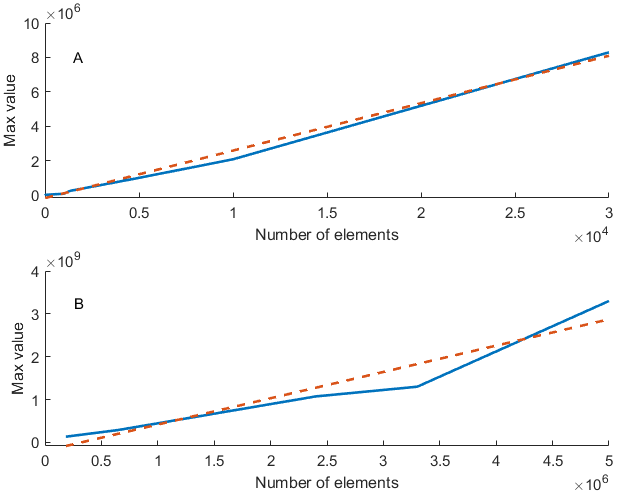}
    \caption{Arrays of differing length and maximum values when the \emph{sqrt} path and \emph{binary search} path have equal duration\\
    \((A)\) Low number of elements, \(R^2 = 0.99275\). \((B)\) High number of elements, \(R^2 = 0.91269\).}
    \label{fig:BestFit}
\end{figure}

When considering this subset of data points, the data forms two distinct clusters: one for a small number of elements (less than about 30,000) and one for a larger number of elements. Figure \ref{fig:BestFit} shows the inflection points where both the \emph{sqrt} and \emph{binary search} paths have equal duration.

For a small number of elements in the input array, the linear best fit is approximately:
\begin{equation}
    275 * numElements - 100000
\end{equation}
For the longer data, the linear fit is:
\begin{equation}
    613 * numElements - 200000000
\end{equation}
When these equations return a value \emph{smaller} than the maximum input value, the \emph{sqrt} path is taken. Otherwise, binary search is selected. These equations perform well for many categories of input arrays, are backed by performance tests, and can readily be adjusted based on new results.

\section{How the results were obtained}
\subsection{Software and hardware}
All results were obtained using MATLAB R2020b on Windows 10, Intel® Core™ i7-9750H CPU @ 2.60GHz, 16 GB RAM. Results were also verified on:
\begin{itemize}
    \item Windows 10, Intel® Xeon® W-2133 CPU @ 3.60GHz, 64 GB RAM
    \item Debian GNU/Linux 10 (buster), Intel® Xeon® Gold 6140 CPU @ 2.30GHz, 51 GiB RAM
    \item macOS Big Sur, Quad-Core Intel® Core i3 @ 3.60GHz, 32 GB RAM
\end{itemize}

\subsection{Reproducibility}
The results on both Windows hardware were mostly consistent within a few percentage points. Linux and Mac fluctuated more but \emph{isprime\_fast} continued to outperform \emph{isprime} strongly for the suite of performance tests in the GitHub repository.

\subsection{Source Code}
The \emph{isprime\_fast} MATLAB source is available on GitHub, \url{ https://github.com/tnear/isprime_fast}.

\section{Results}
\subsection{Scalar performance comparison}
Table \ref{table:PrimePerformance} summarizes the performance of \emph{isprime\_fast}, MATLAB's \emph{isprime}, and Symbolic's \emph{sym/isprime} for prime scalars. The first column, bit-size, is the number of bits of the numeric input. For \emph{isprime} and \emph{isprime\_fast}, double-precision is used for inputs \(\leq\) 2\textsuperscript{53} and unsigned 64-bit integer for inputs > 2\textsuperscript{53} (IEEE 754 double-precision cannot represent odd numbers higher than 2\textsuperscript{53}). The symbolic \emph{sym} data type is used for \emph{sym/isprime}. The timings listed are the average duration of \emph{100 random prime} number inputs with the specified number of bits.

\begin{table}[ht]
\centering
\begin{tabular}{llllll}
Bit-size & \emph{isprime\_fast} (sec) & \emph{isprime} (sec) & \begin{tabular}[c]{@{}l@{}}\emph{isprime\_fast}\\ speedup\end{tabular} & \emph{sym/isprime} (sec) & \begin{tabular}[c]{@{}l@{}}\emph{isprime\_fast}\\ speedup\end{tabular} \\ \hline
4        & 7.7654e-07                 & 4.7186e-06           & 6.07x                                                                  & 0.002166                 & 2789x                                                                  \\
8        & 1.0120e-06                 & 5.7043e-06           & 5.63x                                                                  & 0.00209                  & 2066x                                                                  \\
16       & 1.8106e-06                 & 1.1762e-05           & 6.49x                                                                  & 0.002196                 & 1213x                                                                  \\
24       & 5.8244e-06                 & 4.3816e-05           & 7.52x                                                                  & 0.008325                 & 1429x                                                                  \\
32       & 5.3932e-05                 & 4.1277e-04           & 7.65x                                                                  & 0.0124                   & 230x                                                                   \\
36       & 0.0002828                  & 0.0011006            & 3.89x                                                                  & 0.012055                 & 42.6x                                                                  \\
40       & 0.001766                   & 0.004343             & 2.45x                                                                  & 0.016207                 & 9.17x                                                                  \\
44       & 0.005818                   & 0.01744              & 3.00x                                                                  & 0.015349                 & 2.64x                                                                  \\
48       & 0.02718                    & 0.07891              & 2.90x                                                                  & 0.01777                  & 0.653x                                                                 \\
50       & 0.0326                     & 0.1901               & 5.83x                                                                  & 0.0211                   & 0.648x                                                                 \\
52       & 0.0431                     & 0.3973               & 9.21x                                                                  & 0.0244                   & 0.566x                                                                 \\
56       & 0.0552                     & 1.8212               & 32.9x                                                                  & 0.0226                   & 0.410x                                                                 \\
60       & 0.0755                     & 7.4418               & 98.6x                                                                  & 0.0280                   & 0.370x                                                                 \\
64       & 0.1194                     & 31.8824              & 266x                                                                   & 0.0287                   & 0.234x               
\end{tabular}
\caption{Performance comparison of \emph{prime} scalar inputs of various bit-sizes.}
\label{table:PrimePerformance}
\end{table}

\begin{figure}[ht]
    \centering
    \includegraphics[width=12cm]{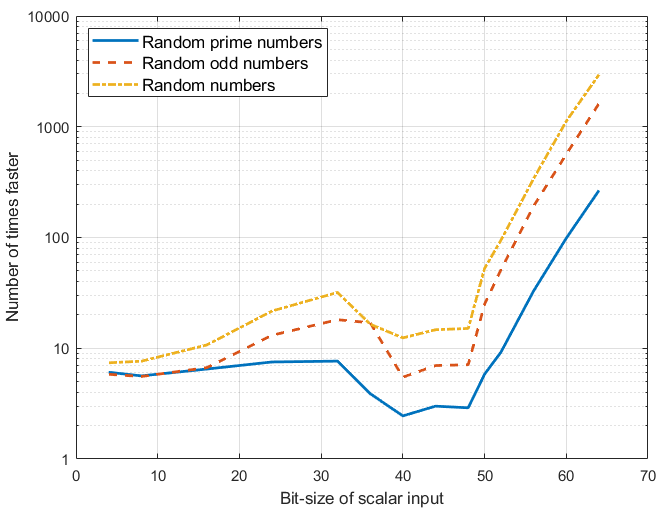}
    \caption{\emph{isprime\_fast} speedup factor compared to \emph{isprime} for various scalar input bit-sizes.}
    \label{fig:Comparison}
\end{figure}

Table \ref{table:PrimePerformance} compares the performance of prime inputs only. Figure \ref{fig:Comparison} summarizes \emph{isprime\_fast}'s performance improvement over \emph{isprime} for prime numbers, odd numbers, and random numbers.

The solid line on the bottom denotes random prime numbers, the dashed line is random odd numbers (complete data in Table \ref{table:OddPerformance}), and the dash-dot line at the top is random numbers (complete data in Table \ref{table:RandomPerformance}).

The improvement of \emph{isprime\_fast} over \emph{isprime} ranges from about 2.4x to 260x faster for random prime numbers depending on the bit-size of the input. For random odd numbers, \emph{isprime\_fast} executes about 5.4x to 1600x faster than \emph{isprime}. And testing random numbers widens the gap further: \emph{isprime\_fast}'s improvement ranges from 7.3x to 2950x faster.

The comparison between \emph{isprime} and \emph{sym/isprime} in Table \ref{table:PrimePerformance} illustrates the importance of data types. \emph{isprime\_fast} is thousands of times faster for small integers: symbolic instructions leave the MATLAB language and incur a noticeable cost for computations which otherwise take microseconds. However, for large primes, \emph{sym/isprime} is 2x to 4x faster than \emph{isprime\_fast} and 9x to 1000x faster than \emph{isprime}. The time that \emph{sym/isprime} takes to convert symbolic MATLAB data into a format recognizable to the C library is more than compensated by that library's faster calculation.

\subsection{Array performance comparison}
Table \ref{table:ArrayPerformance} summarizes the performance comparison results for \emph{isprime\_fast}, \emph{isprime}, and \emph{sym/isprime}  for a variety of array input data types, array lengths, bit-sizes, and data distributions. 

\begin{table}[ht]
\centering
\begin{tabular}{|l|l|l|l|l|l|}
\hline
Input array                                                                                                                                           & \begin{tabular}[c]{@{}l@{}}\emph{isprime\_fast}\\ (sec)\end{tabular} & \begin{tabular}[c]{@{}l@{}}\emph{isprime}\\ (sec)\end{tabular} & \begin{tabular}[c]{@{}l@{}}\emph{isprime\_fast}\\ speedup\end{tabular} & \begin{tabular}[c]{@{}l@{}}\emph{sym/isprime}\\ (sec)\end{tabular} & \begin{tabular}[c]{@{}l@{}}\emph{isprime\_fast}\\ speedup\end{tabular} \\ \hline
1:10000000 (10M)                                                                                                                                      & 0.6294                                                               & 139.50                                                         & 221x                                                                   & 79325*                                                             & 126032x                                                                \\ \hline
\begin{tabular}[c]{@{}l@{}}1609477200:1:(\\ 1609477200 +\\ 3600*24*365)\\ (Unix seconds\\ in the 2021\\ calendar year)\end{tabular}                   & 12.26                                                                & 3286.55                                                        & 268x                                                                   & 26185                                                              & 2136x                                                                  \\ \hline
\begin{tabular}[c]{@{}l@{}}200 random\\ int16 primes\end{tabular}                                                                                     & 5.0952e-05                                                           & 2.2113e-04                                                     & 4.34x                                                                  & 0.0245                                                             & 480x                                                                   \\ \hline
\begin{tabular}[c]{@{}l@{}}1000x1000 matrix\\ of random 32-bit\\ integers\end{tabular}                                                                & 3.927                                                                & 16.647                                                         & 4.24x                                                                  & 529.9                                                              & 134x                                                                   \\ \hline
\begin{tabular}[c]{@{}l@{}}Normal distribution\\ of 1 million primes\\ with mean 2\textsuperscript{16}\end{tabular}                                   & 0.1623                                                               & 5.370                                                          & 33.0x                                                                  & 7614.4                                                             & 46915x                                                                 \\ \hline
\begin{tabular}[c]{@{}l@{}}Normal distribution\\ of 100,000 random\\ odd numbers with\\ mean 2\textsuperscript{32}\end{tabular}                       & 2.689                                                                & 19.603                                                         & 7.28x                                                                  & 144.6                                                              & 53.7x                                                                  \\ \hline
\begin{tabular}[c]{@{}l@{}}Normal distribution\\ of 1,000 random\\ numbers with\\ mean 2\textsuperscript{53}\end{tabular}                             & 5.07                                                                 & 74.32                                                          & 14.6x                                                                  & 0.3370                                                             & 0.0665x                                                                \\ \hline
\begin{tabular}[c]{@{}l@{}}The largest 52-bit\\ prime repeated\\ 1,000 times\end{tabular}                                                             & 58.270                                                               & 172.516                                                        & 2.96x                                                                  & 18.445                                                             & 0.3165x                                                                \\ \hline
\begin{tabular}[c]{@{}l@{}}The 100 largest\\ pseudoprimes \textgreater 2\textsuperscript{53}\\ with factors \textgreater 20M\end{tabular}             & 0.4852                                                               & 7.4524                                                         & 15.3x                                                                  & 0.01160                                                            & 0.0239x                                                                \\ \hline
\begin{tabular}[c]{@{}l@{}}The 100 largest\\ 64-bit primes\end{tabular}                                                                               & 0.1442                                                               & 206.56                                                         & 1432x                                                                  & 0.0284                                                             & 0.197x                                                                 \\ \hline
\begin{tabular}[c]{@{}l@{}}The 5000 largest\\ 64-bit odd integers\end{tabular}                                                                        & 76.201                                                               & 8816.8                                                         & 115.7x                                                                 & 7.951                                                              & 0.1043x                                                                \\ \hline
\end{tabular}
\caption{Performance comparison of non-scalar inputs for \emph{isprime\_fast} \\
\emph{*estimated}}
\label{table:ArrayPerformance}
\end{table}
Based on Table \ref{table:ArrayPerformance}, \emph{isprime\_fast} performs strongly over \emph{isprime} for arrays containing incrementing sequences, prime/odd/composite numbers, normal distributions, random small/medium/large primes, pseudoprimes, and short/medium/long arrays. The performance difference is greatest for large primes and long arrays where \emph{isprime\_fast} often operates hundreds of times faster. The difference is smallest for small integer data types where \emph{isprime\_fast} is only about 2-4x faster. This is due to \emph{isprime} executing faster integer (instead of floating-point) division—\emph{isprime\_fast} has fewer opportunities for optimizations over it.

The metrics for Symbolic's \emph{sym/isprime} reflect the explanation after Table \ref{table:PrimePerformance} about symbolic data. The cost per element to convert to a format recognizable by the external library is massive for long arrays: two of the performance benchmarks indicate that \emph{isprime\_fast} is over 46,000x faster than \emph{sym/isprime}. But for large primes inside short arrays, \emph{sym/isprime} handily beats both \emph{isprime} and \emph{isprime\_fast}.

\section{Discussion}
The largest improvement of \emph{isprime\_fast} over MATLAB's \emph{isprime} is for 64-bit integers where \emph{isprime\_fast} runs 250x to 3000x faster. The modern Miller—Rabin algorithm for 64-bit numbers is tremendously fast compared to \emph{isprime}'s classic prime number sieve followed by prime factor division. One reason for this great contrast is because \emph{isprime}'s implementation predates MATLAB's support for 64-bit integers: the \emph{int64} and \emph{uint64} data types were not supported until 2004 \cite{Moler}. Before then, MATLAB only supported the double-precision data type which has a maximum odd value of \(2^{53} - 1\). This is significant because \emph{isprime} completes in one second for 53-bit numbers but its duration balloons to over 30 seconds for 64-bit integers.

The largest area for improvement of \emph{isprime\_fast} is for scalar inputs in the 40- to 48-bit range. Figure \ref{fig:Comparison} shows a dip in performance for input scalars in this range: \emph{isprime\_fast} only executes two to three times faster than \emph{isprime}. \emph{isprime\_fast} uses a 2/3/5/7-wheel sieve for inputs in this range. For 48-bit integers, this wheel is over 3 million elements long and therefore requires equally as many divisions to determine primality. An alternative algorithm likely could remedy this performance depression.

\section{Conclusions}
MATLAB's \emph{isprime} implementation is stable, effective, and depended on by many mathematics enthusiasts and researchers. \emph{isprime} provides fair performance for small to medium size scalars and short arrays, but less than optimal performance otherwise. \emph{isprime\_fast}, a proposed replacement for \emph{isprime}, improves upon \emph{isprime} for all 2\textsuperscript{64} scalar inputs, ranging from 2.4x to 2950x faster. \emph{isprime\_fast} also surpasses \emph{isprime} for all arrays which were tested, normally ranging from 2x to 1400x faster. \emph{isprime\_fast} efficiently creates wheel sieves, implements the Miller—Rabin test using modular arithmetic strategies, minimizes divisions, and heavily relies on vectorized operations. \emph{isprime\_fast} employs the power of the MATLAB language: it requires no arbitrary-precision arithmetic, C/C++ source code, or external libraries. Using the techniques detailed here, MATLAB's esteemed \emph{isprime} function can be improved considerably.

\subsection{Future research}
\emph{isprime\_fast} still has ample for improvement and additional research. The findings here are just the beginning into the large field of prime number research. \emph{isprime\_fast} uses deterministic Miller–Rabin for large primes. While this is hundreds of times faster than \emph{isprime}'s sieve with trial division for 64-bit integers, it is not the fastest prime number checker. The Baillie–PSW primality test, for example, utilizes one iteration of Miller–Rabin with base 2 followed by a strong Lucas probable prime test \cite{Pomerance}. Implementing the Lucas test without arbitrary-precision arithmetic requires considerable reliance on \emph{ModExp}, \emph{ModMultiply}, and \emph{ModAdd} plus additional modular arithmetic considerations to avoid overflowing 64-bit integers. The results are promising: implemented in MATLAB, Baillie–PSW often runs 1.5x to 2x faster than Miller–Rabin. However, due to the fluctuating nature of Jacobi symbol distributions and the complexity of the Lucas test, the MATLAB implementation still needs additional testing.

\bibliographystyle{abbrv}
\bibliography{references}  

\begin{thebibliography}{10}

\bibitem{Drepper}
U.~Drepper.
\newblock What every programmer should know about memory.
\newblock {\em Red Hat, Inc.}, pages 12,14,56, 2007.

\bibitem{GeeksforGeeks}
GeeksforGeeks.
\newblock How to avoid overflow in modular multiplication?
\newblock
  \url{https://www.geeksforgeeks.org/how-to-avoid-overflow-in-modular-multiplication/},
  2021.

\bibitem{Hare}
I.~Hare.
\newblock Not all cpu operations are created equal. infographics: Operation
  costs in cpu clock cycles.
\newblock
  \url{http://ithare.com/infographics-operation-costs-in-cpu-clock-cycles/},
  2016.

\bibitem{Lynn}
B.~Lynn.
\newblock Number theory – primality tests.
\newblock
  \url{https://crypto.stanford.edu/pbc/notes/numbertheory/millerrabin.html}.

\bibitem{isprime}
MathWorks.
\newblock Determine which array elements are prime.
\newblock \url{https://www.mathworks.com/help/matlab/ref/isprime.html}, 2006.

\bibitem{MATLABData}
MathWorks.
\newblock Matlab data.
\newblock
  \url{https://www.mathworks.com/help/matlab/matlab_external/matlab-data.html},
  2021.

\bibitem{Vectorization}
MathWorks.
\newblock Using vectorization.
\newblock
  \url{https://www.mathworks.com/help/matlab/matlab_prog/vectorization.html},
  2021.

\bibitem{Moler}
C.~Moler and J.~Little.
\newblock A history of matlab.
\newblock {\em Proc. ACM Program. Lang. 4, HOPL, Article 81}, page~24, 2020.

\bibitem{Pomerance}
C.~Pomerance, J.~L. Selfridge, and S.~S. Wagstaff.
\newblock The pseudoprimes to 25*10$^{9}$.
\newblock {\em Mathematics of Computation}, 35:1003--1026, 1980.

\bibitem{Rabin}
M.~Rabin.
\newblock Probabilistic algorithm for testing primality.
\newblock {\em Journal of Number Theory}, pages 128--138, 1980.

\bibitem{Schneier}
B.~Schneier.
\newblock {\em Applied Cryptography: Protocols, Algorithms, and Source Code in
  C}.
\newblock Wiley, 1996.

\bibitem{Sedgewick}
R.~Sedgewick.
\newblock {\em Algorithms in C++}.
\newblock Addison-Wesley, 1992.

\bibitem{Sinclair}
J.~Sinclair.
\newblock Deterministic variants of the miller-rabin primality test.
\newblock \url{https://miller-rabin.appspot.com/}, 2011.

\bibitem{Prime}
N.~J.~A. Sloane.
\newblock a(0) = 1; for n > 0, a(n) = (prime(n)-1)*a(n-1).
\newblock \url{https://oeis.org/A005867}.

\bibitem{Primorial}
N.~J.~A. Sloane.
\newblock Primorial numbers: product of first n primes.
\newblock \url{https://oeis.org/A002110}.

\end{thebibliography}

\section{Appendix tables}
\begin{table}[ht]
\centering
\begin{tabular}{llllll}
Bit-size & \emph{isprime\_fast} (sec) & \emph{isprime} (sec) & \begin{tabular}[c]{@{}l@{}}\emph{isprime\_fast}\\ speedup\end{tabular} & \emph{sym/isprime} (sec) & \begin{tabular}[c]{@{}l@{}}\emph{isprime\_fast}\\ speedup\end{tabular} \\ \hline
4        & 7.9037e-07                 & 4.5998e-06           & 5.82x                                                                  & 0.00211                  & 2668x                                                                  \\
8        & 9.8097e-07                 & 5.4535e-06           & 5.56x                                                                  & 0.0020                   & 2084x                                                                  \\
16       & 1.4468e-06                 & 9.6701e-06           & 6.68x                                                                  & 0.0020                   & 1409x                                                                  \\
24       & 2.9905e-06                 & 3.9433e-05           & 13.1x                                                                  & 0.00299                  & 1000x                                                                  \\
32       & 1.9783e-05                 & 3.5878e-04           & 18.1x                                                                  & 0.00298                  & 150x                                                                   \\
36       & 2.3884e-05                 & 4.0707e-04           & 17.0x                                                                  & 0.002895                 & 121x                                                                   \\
40       & 7.2903e-04                 & 0.0040               & 5.48x                                                                  & 0.0027                   & 3.49x                                                                  \\
44       & 0.002690                   & 0.01878              & 6.98x                                                                  & 0.003328                 & 1.23x                                                                  \\
48       & 0.01062                    & 0.07565              & 7.1x                                                                   & 0.002771                 & 0.260x                                                                 \\
50       & 0.00636                    & 0.1616               & 25x                                                                    & 0.00279                  & 0.439x                                                                 \\
52       & 0.007706                   & 0.39378              & 51x                                                                    & 0.002832                 & 0.367x                                                                 \\
56       & 0.0095                     & 1.8229               & 192x                                                                   & 0.0044                   & 0.463x                                                                 \\
60       & 0.01316                    & 7.4327               & 564x                                                                   & 0.003645                 & 0.277x                                                                 \\
64       & 0.0196                     & 31.7576              & 1620x                                                                  & 0.0034                   & 0.174x                    
\end{tabular}
\caption{Performance comparison of random \emph{odd} scalar inputs of various bit-sizes.}
\label{table:OddPerformance}
\end{table}

\begin{table}[ht]
\centering
\begin{tabular}{llllll}
Bit-size & \emph{isprime\_fast} (sec) & \emph{isprime} (sec) & \begin{tabular}[c]{@{}l@{}}\emph{isprime\_fast}\\ speedup\end{tabular} & \emph{sym/isprime} (sec) & \begin{tabular}[c]{@{}l@{}}\emph{isprime\_fast}\\ speedup\end{tabular} \\ \hline
4        & 5.4788e-07                 & 4.0505e-06           & 7.39x                                                                  & 0.002046                 & 3734x                                                                  \\
8        & 7.1275e-07                 & 5.4507e-06           & 7.64x                                                                  & 0.002103                 & 2950x                                                                  \\
16       & 9.0833e-07                 & 9.7864e-06           & 10.7x                                                                  & 0.002008                 & 2210x                                                                  \\
24       & 1.8587e-06                 & 4.0482e-05           & 21.7x                                                                  & 0.002289                 & 1231x                                                                  \\
32       & 1.1610e-05                 & 3.7037e-04           & 31.9x                                                                  & 0.002584                 & 222x                                                                   \\
36       & 6.9472e-05                 & 0.001146             & 16.5x                                                                  & 0.002507                 & 36.0x                                                                  \\
40       & 3.3226e-04                 & 0.004121             & 12.4x                                                                  & 0.002913                 & 8.76x                                                                  \\
44       & 0.001256                   & 0.01854              & 14.7x                                                                  & 0.002359                 & 1.87x                                                                  \\
48       & 0.004622                   & 0.06985              & 15.1x                                                                  & 0.002308                 & 0.4993x                                                                \\
50       & 0.003334                   & 0.1755               & 52.6x                                                                  & 0.002601                 & 0.7801x                                                                \\
52       & 0.003863                   & 0.3647               & 94.4x                                                                  & 0.002498                 & 0.6467x                                                                \\
56       & 0.005152                   & 1.7537               & 340x                                                                   & 0.003188                 & 0.6189x                                                                \\
60       & 0.006451                   & 7.2382               & 1122x                                                                  & 0.002881                 & 0.4466x                                                                \\
64       & 0.01126                    & 33.2620              & 2953x                                                                  & 0.002531                 & 0.2247x                   
\end{tabular}
\caption{Performance comparison of \emph{random} scalar inputs of various bit-sizes.}
\label{table:RandomPerformance}
\end{table}

\end{document}